\documentclass[useAMS,usenatbib]{mn2e} 
\usepackage{times,psfig,epsfig} 
\newcommand{\mnras}{MNRAS}
\newcommand{\aap}{A\&A}

\newcommand{\apj}{ApJ}

\newcommand{\apjl}{ApJ}
\newcommand{\physrep}{Phys. Rep.}
\newcommand{\araa}{ARA\&A}

\title[SZ profiles and scaling relations: modelling effects and 
observational biases]{Sunyaev-Zel'dovich 
profiles and scaling relations: modelling effects and 
observational biases}
\author[A. Bonaldi et al.]
{A.  Bonaldi$^{1,2}$, G. Tormen$^{2}$, 
K. Dolag$^{3}$ and  L. Moscardini$^{4,5}$ \\
$^{1}$ INAF-Osservatorio Astronomico di Padova, 
vicolo dell'Osservatorio 5, I-35122 Padova, Italy
(anna.bonaldi@oapd.inaf.it)\\
$^{2}$ Dipartimento di Astronomia, Universit\`a di
Padova, vicolo dell'Osservatorio 2, I-35122 Padova, Italy
(giuseppe.tormen@unipd.it)\\
$^{3}$ Max-Planck Institut fuer Astrophysik,
Karl-Schwarzschild Strasse 1, D-85748 Garching, Germany
(kdolag@mpa-garching.mpg.de)\\
$^{4}$ Dipartimento di Astronomia, Universit\`a di Bologna,
via Ranzani 1, I-40127 Bologna, Italy
(lauro.moscardini@unibo.it)\\
$^{5}$ INFN, Sezione di Bologna, viale Berti Pichat 6/2,
I-40127 Bologna, Italy
}

\begin{document}

\date{Received April 2007; accepted by MNRAS 2007 April 19; in original form August 2006}

\pagerange{\pageref{firstpage}--\pageref{lastpage}} \pubyear{2006}

\maketitle

\label{firstpage}

%##############################################################################
%###################### Abstract ##############################################
%##############################################################################

\begin{abstract} 

We use high-resolution hydrodynamic re-simulations to investigate the properties 
of the thermal Sunyaev-Zel'dovich (SZ) effect from galaxy clusters. We compare 
results obtained using different physical models for the intracluster medium (ICM), 
and show how they modify the SZ emission in terms of cluster profiles and scaling
relations. We also produce realistic mock observations to verify whether the results 
from hydrodynamic simulations can be confirmed. 
We find that SZ profiles depend marginally on the modelled physical processes, while 
they exhibit a strong dependence on cluster mass.  The central and total SZ emission 
strongly correlate with the cluster X-ray luminosity and temperature. 
The logarithmic slopes of these scaling relations differ from the self-similar 
predictions by less than 0.2; the normalization of the relations is lower for 
simulations including radiative cooling. The observational test suggests that SZ 
cluster profiles are unlikely to be able to probe the ICM physics. The total SZ 
decrement appears to be an observable much more robust than the central intensity, and we suggest using the former to investigate scaling relations.

\end{abstract} 


%##############################################################################
%########################## Introduction ######################################
%##############################################################################

\section{Introduction}  

Galaxy clusters are ideal probes for studies of large scale structures.
They have typical masses of order $10^{14}-10^{15} M_{\odot}$; within
a radius of a few Mpc they contain hundreds of galaxies orbiting in a
gravitational potential well due primarily to dark matter (DM).
They are also filled with ICM: hot ionized gas, typically at a temperature 
of 1-15 keV, that can be observed via its bremsstrahlung emission in the 
soft X-ray band, and via the Sunyaev-Zel'dovich (SZ) effect in the millimetric
band.

Several observational evidences demonstrate that the picture of an ICM
in hydrostatic equilibrium in the DM potential well of the cluster is
an over-simplification: above all, scaling relations between various
cluster properties, like mass, temperature and X-ray luminosity,
differ from the self-similar predictions \citep[see, e.g., the recent
reviews by][and the references therein]{rosati2002,voit2005,borgani2006}.  In
order to explain the ICM properties, many models have been developed
so far, which consider the effect of radiative cooling and of
non-gravitational heating (mainly feedback from supernovae and
AGNs). As the complexity in the physical description of the ICM
increased, hydrodynamic simulations have also improved to include such
processes, and have become invaluable tools to compare theoretical
predictions with current data and to forecast the performance of future experiments.

On the observational side, important improvements are expected for the
detection of the SZ effect in clusters thanks to an upcoming new generation 
of suitable instruments, like
AMI\footnote{\emph{http://www.mrao.cam.ac.uk/telescopes/ami/}},
SPT\footnote{\emph{http://spt.uchicago.edu/}},
ACT\footnote{\emph{http://www.hep.upenn.edu/act/}},
SZA\footnote{\emph{http://astro.uchicago.edu/sza/}},
AMiBA\footnote{\emph{http://amiba.asiaa.sinica.edu.tw/}},
APEX\footnote{\emph{http://bolo.berkeley.edu/apexsz/}},
and {\sc Planck}\footnote{\emph{http://www.rssd.esa.int/Planck/}}.
In order to correctly extract and utilize the information contained in
these data it will be mandatory to keep under control both the dependence
of the SZ signal on the ICM physics, and the possible biases induced
on the intrinsic signal by the details of the observational process.

In the present work we investigate how different choices for the
physical modelling of the ICM affect the SZ radial profiles and
the scaling relations between the SZ flux and other intrinsic
properties of clusters.  To this end, we use a set of
high-resolution hydrodynamical simulations, where clusters were 
re-simulated including different physical processes, from radiative
cooling, to star formation, energy feedback from supernovae, and
thermal conduction.

After this analysis, we adopt the instrumental characteristics of
the AMI interferometer and simulate the observational process for a
few clusters in our sample: this enable us to discuss possible biases
introduced in our results by a realistic observation process.

The present paper is organized as follows. After briefly introducing in
Sect. \ref{sec:uno} the SZ effect and the main related quantities, in
Sect. \ref{sec:due} we describe the general characteristics of the
hydrodynamic simulations used in our study.  In Sect. \ref{sec:tre} we
obtain fitting formulas for the cluster SZ profiles, while in
Sect. \ref{sec:quattro} we discuss the effects of various physical
processes on the scaling relations between some global properties commonly
used to describe galaxy clusters.  In Sect. \ref{sec:cinque} we perform
simulated observations assuming the characteristics of the AMI
instrument, and discuss the robustness of our results.  Finally in
Sect. \ref{sec:conclu} we draw our conclusions.

%##############################################################################
%################### The Sunyaev-Zel'dovich effect#############################
%##############################################################################

\section{The Sunyaev-Zel'dovich effect}\label{sec:uno} 

The thermal SZ effect \citep{sunyaev1972} is a distorsion in the
cosmic microwave background (CMB) spectrum due to inverse Compton
scattering of CMB photons by hot ionized gas particles
\citep[for recent reviews see, e.g.,][]{birkinshaw1999,carlstrom2002,rephaeli2005}. 
Its intensity is given by the Comptonization parameter $y$, defined as:  
\begin{equation} 
y\equiv  \frac{k_B\sigma_T}{m_{e}c^2} \int n_{e} T_{e} dl\ , 
\label{eq:y} 	
\end{equation} 
where $k_B$ is the Boltzmann constant, $\sigma_T$ is the Thomson cross
section, $c$ is the speed of light, and $n_e$, $T_e$ and $m_e$ are the
electronic number density, temperature and rest mass, respectively. As
evident from Eq.(\ref{eq:y}), the $y$-parameter is directly
proportional to the cluster pressure $P$ integrated along the line of
sight.  The change in the CMB intensity at the frequency $\nu$
corresponding to a certain value of the Comptonization parameter is:
\begin{equation} 
\Delta I_{\rm CMB}=\frac{2(k_BT)^3}{(hc)^2}g(x)\ y\ ,
\end{equation} 
where $h$ is the Planck constant.  The frequency dependence is
given by the spectral shape $g(x)$:
\begin{equation} 
g(x)=\frac{x^4 e^x}{(e^x -1)^2}\left(x\frac{e^x +1}{e^x -1}-4\right)\ ,
\end{equation}
where $x\equiv h\nu/k_BT_{\rm CMB}$.
  
In galaxy clusters the Comptonization parameter $y$ typically reaches 
values of order $10^{-6}$ on arc-minute scales, corresponding to
fluctuations in the CMB temperature of few $\mu K$. As a consequence,
observing galaxy clusters through the SZ effect requires extremely good 
sensitivity, in order to obtain sufficiently high signal-to-noise ratios.
The required sensitivity and sky coverage will be achieved in the near 
future, thanks to new dedicated instruments.  Nevertheless, 
the resolution of these instruments will be not enough to detect
the wealth of details which can be seen in numerical hydrodynamical 
simulations: substructures, shocks, etc. For this reason, in order to 
be meaningful and helpful, all analyses of the SZ effect based on
simulations must deal with the actual instrumental capabilities: this
is one of the main goals of this paper, as discussed below.

%##############################################################################
%################### simulated clusters #######################################
%##############################################################################

\section{Properties of the simulated clusters}\label{sec:due} 
 
The hydrodynamic simulations used in the present work were carried out using 
{\tt GADGET-2} \citep{springel2005}, a new version of the parallel Tree+SPH
(Smoothed Particle Hydrodynamics) code {\tt GADGET} \citep{springel2001}.  
{\tt GADGET-2} includes an entropy-conserving formulation of SPH 
\citep{springel2002}, radiative cooling, heating by an UV background, and a 
treatment of star formation and feedback from galactic winds powered by 
supernova explosions.  The latter is based on a subresolution model for the 
multi-phase structure of the interstellar medium, as described in 
\cite{springel2003}.  The SPH implementation of thermal conduction in 
{\tt GADGET-2}, which is both stable and manifestly conserves thermal energy - 
even when individual and adaptive time-steps are used - was described in
\cite{jubelgas2004}.  This implementation assumes an isotropic effective 
conductivity parameterized as a fixed fraction of the Spitzer rate. 
It also accounts for saturation, which can become relevant for low-density gas.

\subsection{The sample of re-simulated clusters} 

\begin{table} 
\caption{Different physical processes included in our re-simulations} 
\begin{tabular}{|c|c|} 
\hline 
simulation name& involved physical processes\\ 
\hline 
\emph{ovisc} & non-radiative gas-dynamics with standard viscosity\\ 
\emph{lvisc} & like \emph{ovisc}, but with low viscosity\\ 
\emph{csf}   & cooling, star formation and feedback\\ 
\emph{csfc}  & like \emph{csf} plus thermal conduction\\ 
\hline 
\label{tab:uno} 
\end{tabular} 
\end{table} 
 
\begin{table*}
\caption{The sample of simulated clusters.
For each object we report name, virial mass $M_{\rm vir}$, 
virial radius $R_{\rm vir}$, mass-weighted temperature $T$,
X-ray luminosity $L_X$ in the [0.1-10 keV] band, and central value
$y_0$ of the Comptonization parameter.  Clusters are divided in two
samples depending on their mass: low-mass (LM) and high-mass (HM).}
\label{tab:unobis} 
\begin{tabular}{|c|c|c|c|c|c|} 
\hline 
LM sample& $M_{\rm vir}$ [$M_{\odot}/h$] & $R_{\rm vir}$ [kpc/$h$]
& $T$ [keV] & $L_X$ [erg/s] & $y_0$ \\ 
\hline 
\emph{g676} &$1.1\times10^{14}$& 983& 1.1&$3.2\times10^{44}$ & 
$3.6\times 10^{-5}$\\ 
\emph{g3344}&$1.1\times10^{14}$&1002& 1.4&$2.2\times10^{44}$ & 
$2.9\times 10^{-5}$\\
\emph{g6212}&$1.1\times10^{14}$&1000& 1.3&$3.0\times10^{44}$ & 
$4.4\times 10^{-5}$\\ 
\emph{g1542}&$1.1\times10^{14}$& 982& 1.3&$3.0\times10^{44}$ & 
$4.2\times 10^{-5}$\\ 
\emph{g1b}  &$4.5\times10^{14}$&1585& 2.6&$4.2\times10^{44}$ & 
$5.0\times 10^{-5}$\\ 
\emph{g1c}  &$1.9\times10^{14}$&1181& 1.4&$3.0\times10^{44}$ & 
$4.3\times 10^{-5}$\\ 
\emph{g1d}  &$1.5\times10^{14}$&1098& 1.2&$1.0\times10^{44}$ & 
$1.4\times 10^{-5}$\\
\hline 
HM sample & $M_{\rm vir}$ [$M_{\odot}/h$] & $R_{\rm vir}$ [kpc/$h$]
& $T$ [keV] & $L_X$ [erg/s] & $y_0$ \\ 
\hline 
\emph{g8a}  & $2.4\times10^{15}$&2758&10.6&$8.1\times10^{45}$ & 
$7.4\times 10^{-4}$\\ 
\emph{g1a}  & $1.5\times10^{15}$&2358& 7.0&$4.8\times10^{45}$ & 
$4.3\times 10^{-4}$\\ 
\emph{g51a} & $1.3\times10^{15}$&2276& 6.2&$3.2\times10^{45}$ & 
$3.5\times 10^{-4}$\\ 
\emph{g72a} & $1.4\times10^{15}$&2299& 5.8&$1.8\times10^{45}$ & 
$1.9\times 10^{-4}$\\
\hline 
\end{tabular} 
\end{table*} 

The set of simulated galaxy clusters used in the following analysis
includes 11 objects, which we divide in two samples: seven low-mass
clusters, with virial mass $M_{\rm vir}\simeq 10^{14} h^{-1}M_{\odot}$
(LM sample), and four high-mass systems with $M_{\rm vir} =
(1.3-2.3)\times 10^{15} h^{-1} M_{\odot}$ (HM sample).  The cluster
regions were extracted from a (DM-only) cosmological simulation
\citep[see][]{yoshida2001} with box-size $479\,h^{-1}\,{\rm Mpc}$ of a
flat $\Lambda$CDM model with matter density parameter $\Omega_{\rm
0m}=0.3$, cosmological constant density parameter
$\Omega_\Lambda=0.7$, dimensionless Hubble parameter $h=0.7$, power spectrum
normalization $\sigma_8=0.9$ and baryon density parameter $\Omega_{\rm
b}=0.04$.  Using the ``Zoomed Initial Conditions'' (ZIC) technique
\citep{tormen1997}, these regions were re-simulated at higher
mass and force resolution by populating their Lagrangian volume in
the initial conditions with a larger number of particles, while
appropriately adding high-frequency modes. At the same time, the
large-scale tidal field of the cosmological environment is correctly
described by using low-resolution particles.  The unperturbed particle
distribution (before displacements) is of glass-type
\citep{white1996}.  Gas was added only in the high-resolution region
by splitting each original particle into a gas and a DM one. Thereby,
the gas and the DM particles were displaced by half the original mean
interparticle distance, so that the centre-of-mass and the momentum
were conserved. Each gas particle has mass $m_{\rm gas}=1.7\times
10^8\,h^{-1}\,M_{\odot}$, while the mass of the DM particles is
$m_{\rm DM}=1.13\times 10^9\,h^{-1}\,M_{\odot}$; thus, the total
number of particles inside our simulated clusters is between
$2\times10^5$ and $4\times10^6$, depending on their final mass.  For
all simulations, the gravitational softening length was kept fixed at
$\epsilon=30\,h^{-1}\,\mathrm{kpc}$ comoving (Plummer-equivalent) for
$z>5$, and was switched to a physical softening length of
$\epsilon=5\,h^{-1}\,\mathrm{kpc}$ at $z=5$.

Each object of the cluster set was re-simulated four times,
including a different set of physical processes for the gas component
(see the list in Table \ref{tab:uno}).  Both simulations referred as
\emph{ovisc} and \emph{lvisc} follow non-radiative gas-dynamics: 
for \emph{ovisc} a standard viscosity scheme is adopted, while
\emph{lvisc} runs were carried out using a modified artificial
viscosity scheme suggested by \cite{morris1997}, where each particle
evolves in time its individual viscosity parameter.  Whereas in this
scheme shocks are as well captured as in the standard scheme, regions
with no shocks do not suffer for residual non-vanishing artificial
viscosity. Therefore, turbulence driven by fluid instabilities can be
much better resolved by the new scheme and, as a result of this,
galaxy clusters can build up a sufficient level of turbulence-powered
instabilities along the surfaces of the large-scale velocity structure
present in cosmic structure formation \citep{dolag2005}.  In
simulations which include star formation and feedback (\emph{csf} and
\emph{csfc}), the supernovae efficiency in powering galactic winds was
set to 50 per cent, which corresponds to a wind speed of about 340 km
s$^{-1}$.  For the \emph{csfc} simulations, which also include the
effect of thermal conduction, we assume a fixed fraction 1/3 of the
Spitzer rate.  A more detailed description of the properties of
simulated galaxy clusters including thermal conduction can be found in
\cite{dolag2004}.

Table \ref{tab:unobis} summarizes the main properties of the 11
clusters, obtained using the \emph{ovisc} simulation.  In particular,
for each object we list its virial mass $M_{\rm vir}$, virial radius 
$R_{\rm vir}$, mass-weighted temperature $T$, X-ray luminosity $L_X$ 
in the [0.1-10 keV] band, and central value $y_0$ of the Comptonization 
parameter.  Notice that both $T$ and $L_X$ are computed inside $R_{\rm vir}$, 
which was defined using the overdensity threshold dictated by the spherical 
top-hat model \citep[see, e.g.,][]{eke1996}. 
The same properties in simulations with a different ICM modelling show 
discrepancies at most of few per cent for virial masses and
mass-weighted temperatures. Conversely, typical X-ray luminosities,
when compared to the \emph{ovisc} runs, are reduced on average by a
factor of 2 in the \emph{lvisc} simulations and by a factor of 5 in
both the \emph{csf} and \emph{cfsc} runs.

%##############################################################################
%################### Fitting ##################################################
%##############################################################################

\section{Fitting radial profiles}\label{sec:tre} 
 
In this section we discuss the procedure adopted to find fitting relations
for the radial profiles of the SZ maps of our clusters. To produce 
two-dimensional (2D) SZ maps we followed a standard procedure which employs
the SPH kernel to distribute the desired quantities (in this case the 
Comptonization parameter $y$), on a grid. Details on the method may be found 
in \cite{dolag2006} and \cite{roncarelli2006}. In particular, we considered 
the simulation outputs at $z=0$, and for each cluster we obtained three SZ 
maps, corresponding to projections along the cartesian axes. The SZ maps, 
4 Mpc on a side, were integrated over 8 Mpc along the line of sight: we 
checked that such integration is enough to enclose most of the cluster 
signal, so that no artificial cutoff is introduced. 

\begin{figure*} 
\begin{center}
 \includegraphics[width=0.95\textwidth]{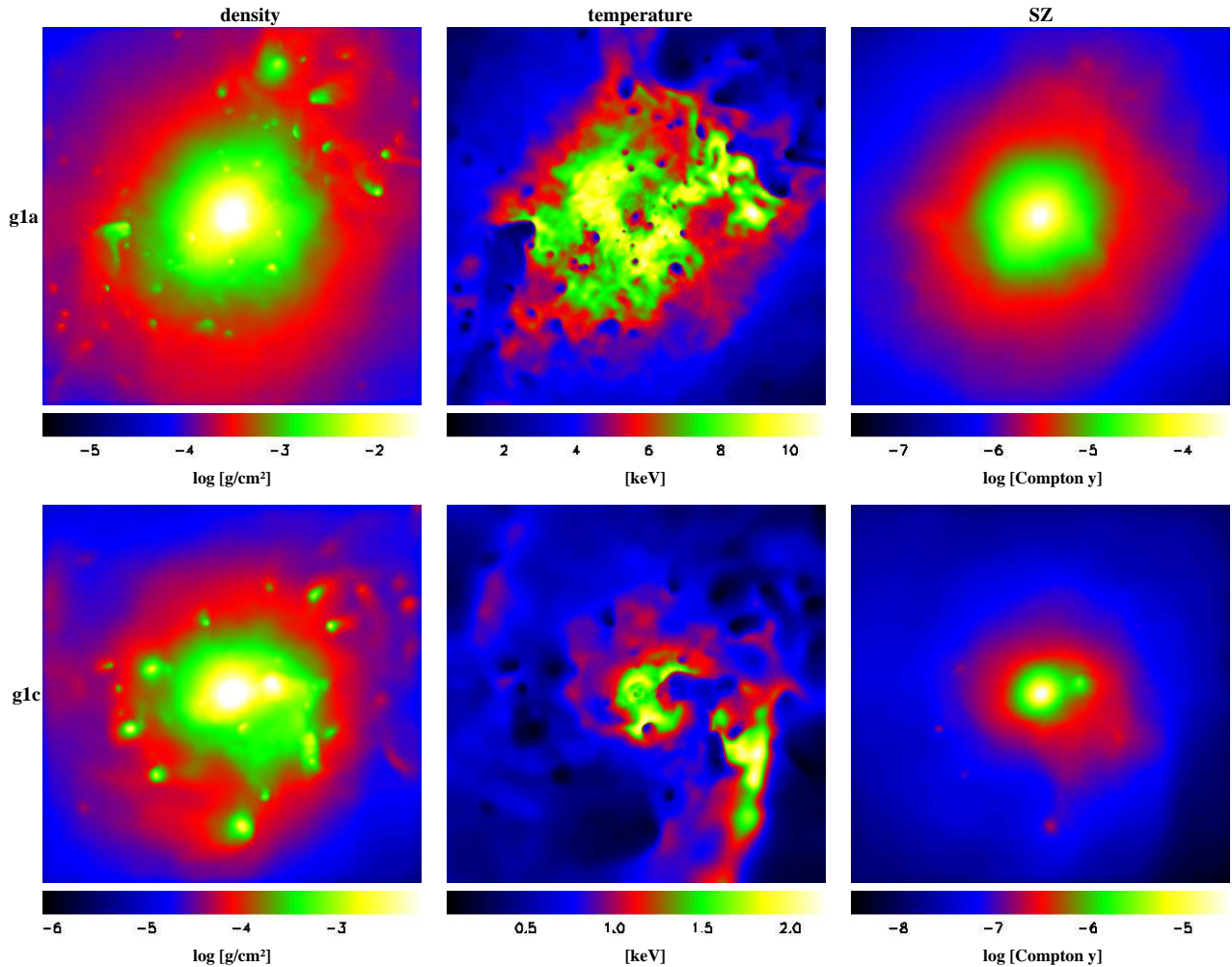} 
\caption{
Projected maps of gas density (left column), mass-weighted temperature
(central column) and Comptonization parameter $y$ (right column) for
two galaxy clusters: \emph{g1a} (top panels) and \emph{g1c}
(bottom panels), both taken at $z=0$ and considering the non-radiative
simulation \emph{ovisc}.  The displayed region is a square with side 4
Mpc. The colour scale is shown at the bottom of each panel.}
\label{fig:zero} 
\end{center} 
\end{figure*}

Examples of SZ maps are shown on the right column of Fig.\ref{fig:zero}, 
where we present two systems, \emph{g1a} (upper panels) and \emph{g1c}
(bottom panels), representative of the HM and LM samples,
respectively.  In both cases we show the output of the non-radiative 
simulation \emph{ovisc}.  For completeness we also display the corresponding 
2D maps for the gas density (left column) and for the mass-weighted 
temperature (central column).

The radial profiles obtained from the SZ maps are strongly 
dependent on the cluster mass: for this reason we decided to obtain 
fitting profiles for the LM and HM samples separately.
In order to obtain best-fitting relations for the profiles of the
Comptonization parameter $y$, we started from three-dimensional (3D)
data, and built a model for the SZ emission directly linked to the 3D
cluster shape. In particular, we obtained the 3D pressure profile for
each cluster of our sample and fit it using a simple analytical formula, 
which we then integrated along the line of sight. As we are going to 
show, the resulting expression, evaluated at the correct integration 
limits, provides a good fit for the SZ radial profiles.

The 3D pressure profiles of each cluster were obtained by calculating
the mean value of $\varrho T$ for particles in spherical shells of
equal logarithmic width out to $4 R_{\rm vir}$.  In order to extract a
mean profile for each mass bin, and for each set-up of the physical
processes, we need to appropriately normalize each profile, obtaining
dimensionless quantities. Assuming a self-similar model for the density
profiles, differences in pressure are due to differences in
temperature only. We found that the resulting pressure profiles for
our simulated clusters, however, do not properly superimpose when
normalized by their temperature, indicating that also density profiles
show non-negligible deviations from a self-similar behaviour. For this
reason we preferred to use an empirical normalization, adopting the value
of the pressure integrated up to a given fixed radius. We chose
$R_{500}$, the radius at which the mean overdensity of the cluster is
500 times the critical density of the universe (roughly corresponding
to 60 per cent of $R_{\rm vir}$). The normalization factor obtained in 
this way for each cluster was also used to obtain the mean SZ profiles.
 
After obtaining a mean 3D profile for each cluster subsample, we looked
for a fitting relation that is also integrable along the line of
sight.  We chose the following expression for the 3D fit in spherical 
coordinates $(R,\theta,\phi)$:
\begin{equation} 
F(R) = I_{3D} (R+r_p)^b + K \ , \\
\label{3dfit}
\end{equation} 
where $R=\sqrt {x^2+y^2+z^2}$ is the 3D clustercentric distance (in
units of the virial radius $R_{\rm vir}$) and $I_{3D}$ is the
intensity normalization factor. The free parameters used for the fit
are three: $r_p$ (also given in units of $R_{\rm vir}$), $b$ and
$K$.

In order to obtain the SZ map fitting relation we rewrite
Eq.(\ref{3dfit}) in cylindrical coordinates $(r,\theta,z)$ and
integrate along $z$, obtaining:
\begin{eqnarray} 
f(r)&=&I_{2D} \int_{l_1}^{l_2}F(r,z)dz =\nonumber \\ 
&=&I_{2D}\,\Big[\left(r_{p}+\sqrt{r^2+z^2}
\right)^{(b-1)}+Kz\Big]_{z=l_1}^{z=l2}\ ,
\label{2dfit}
\end{eqnarray} 
where $r=\sqrt {x^2+y^2}$ is the 2D clustercentric distance (again in
units of the virial radius $R_{\rm vir}$) and $l_1$, $l_2$ are the
integration limits adopted.  From a theoretical point of view,
integration along the line of sight should be performed from the
centre to infinity. For practical reasons, however, the outer limit 
can be safely set to a radius above which the intensity does not
significantly contribute to the result: in our case we choose
$l_2=2\times R_{\rm vir}$.  As for the inner limit, the 3D profile
in the simulations can be robustly computed only down to radii
enclosing at least of order of 100 particles, so our fitting profile 
for $R \rightarrow 0$ is actually an extrapolation. In particular, 
the contribution from the cluster centre to the integral along the
$z$-axis leads to errors in the SZ profile. However, if we give up 
integrating down to $z \rightarrow 0$, we can minimize the contribution 
from $R \rightarrow 0$ and get a correct integrated profile. For this
reason we used $l_1=0.015 R_{\rm vir}$ and we accounted for the flux 
underestimation in the factor $I_{2D}$.

\begin{figure*} 
\begin{center} 
\includegraphics[width=8cm,height=7.3cm,angle=90]{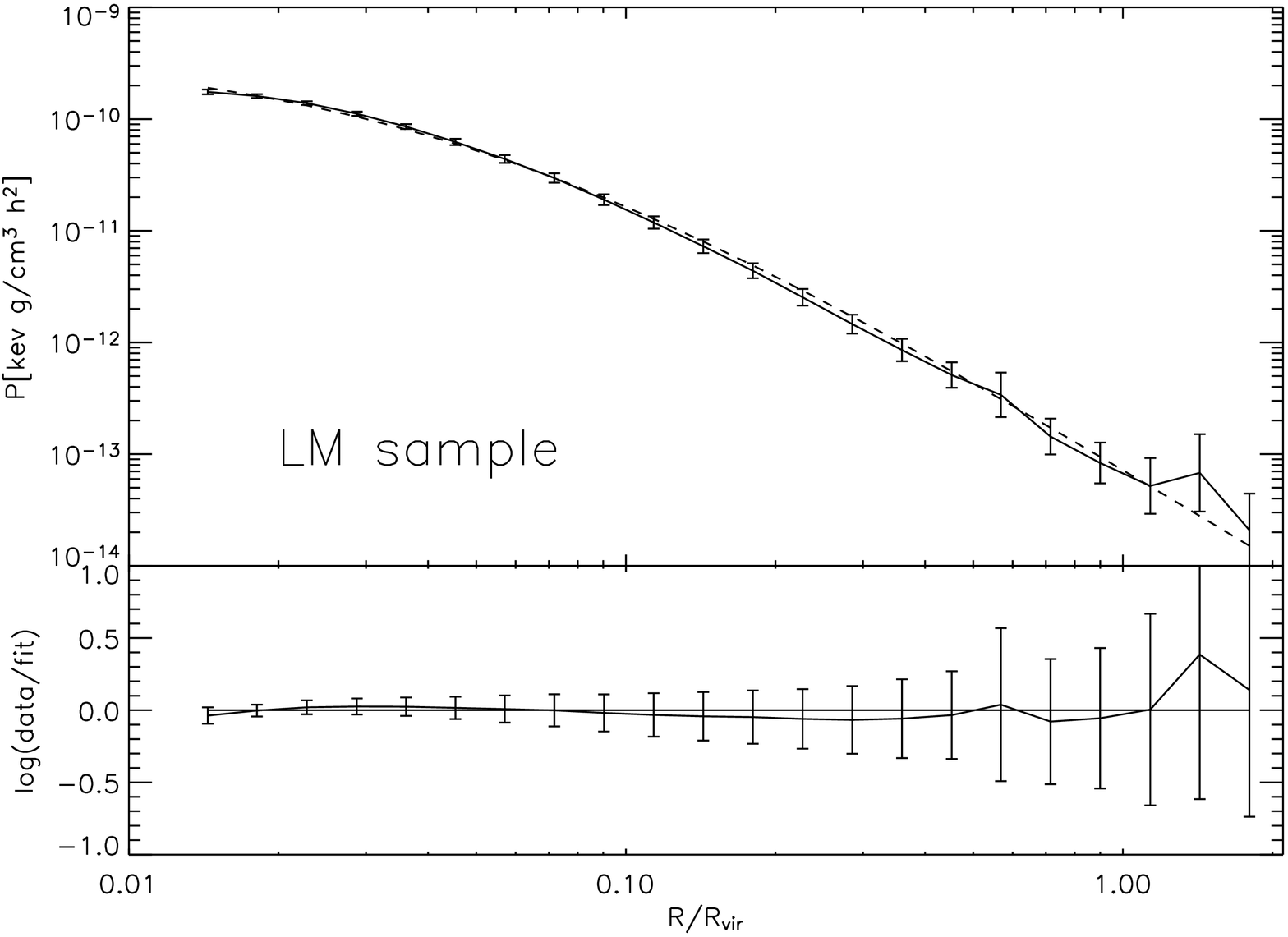} 
\includegraphics[width=8cm,height=7.3cm,angle=90]{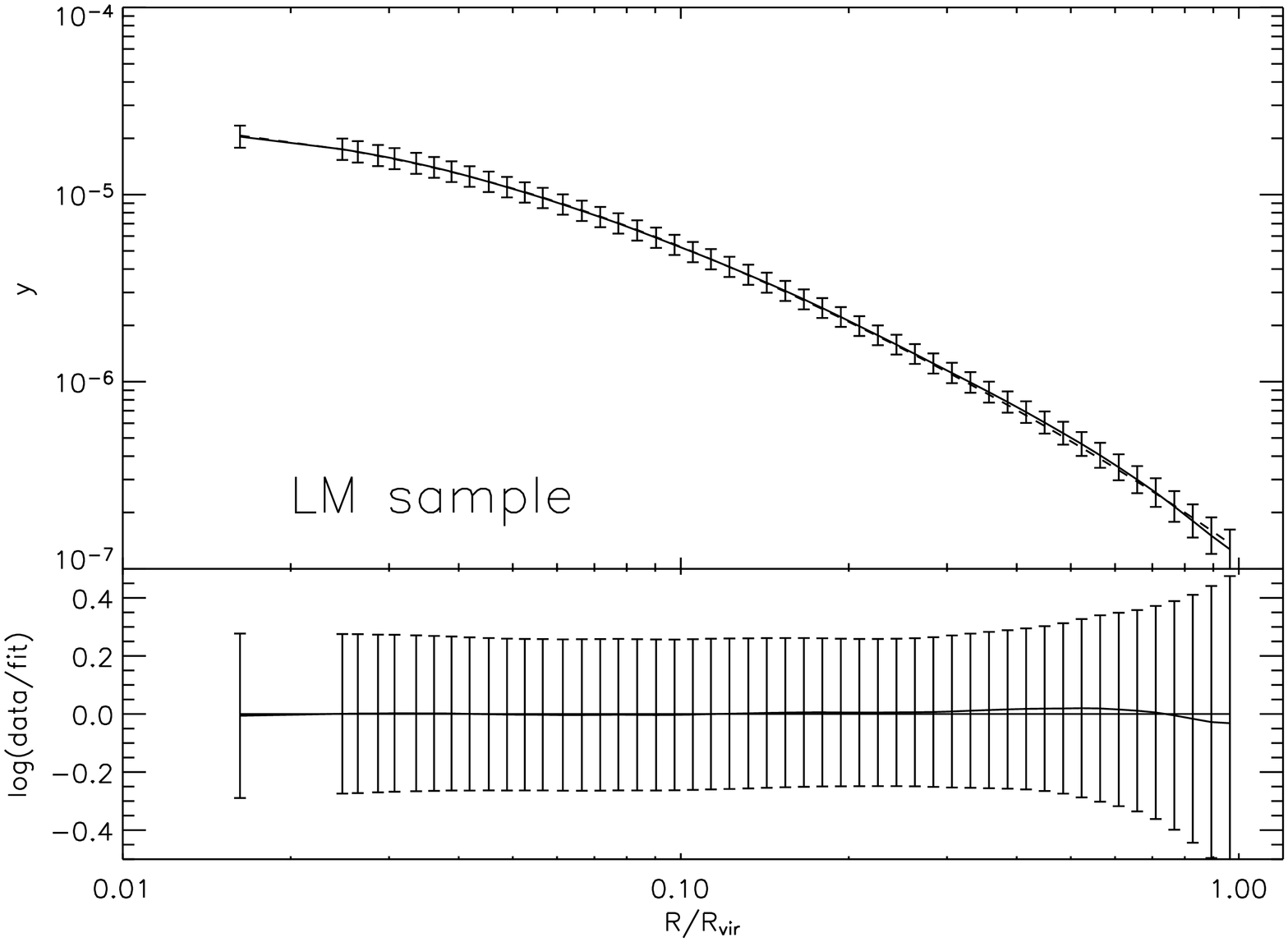} 
\includegraphics[width=8cm,height=7.3cm,angle=90]{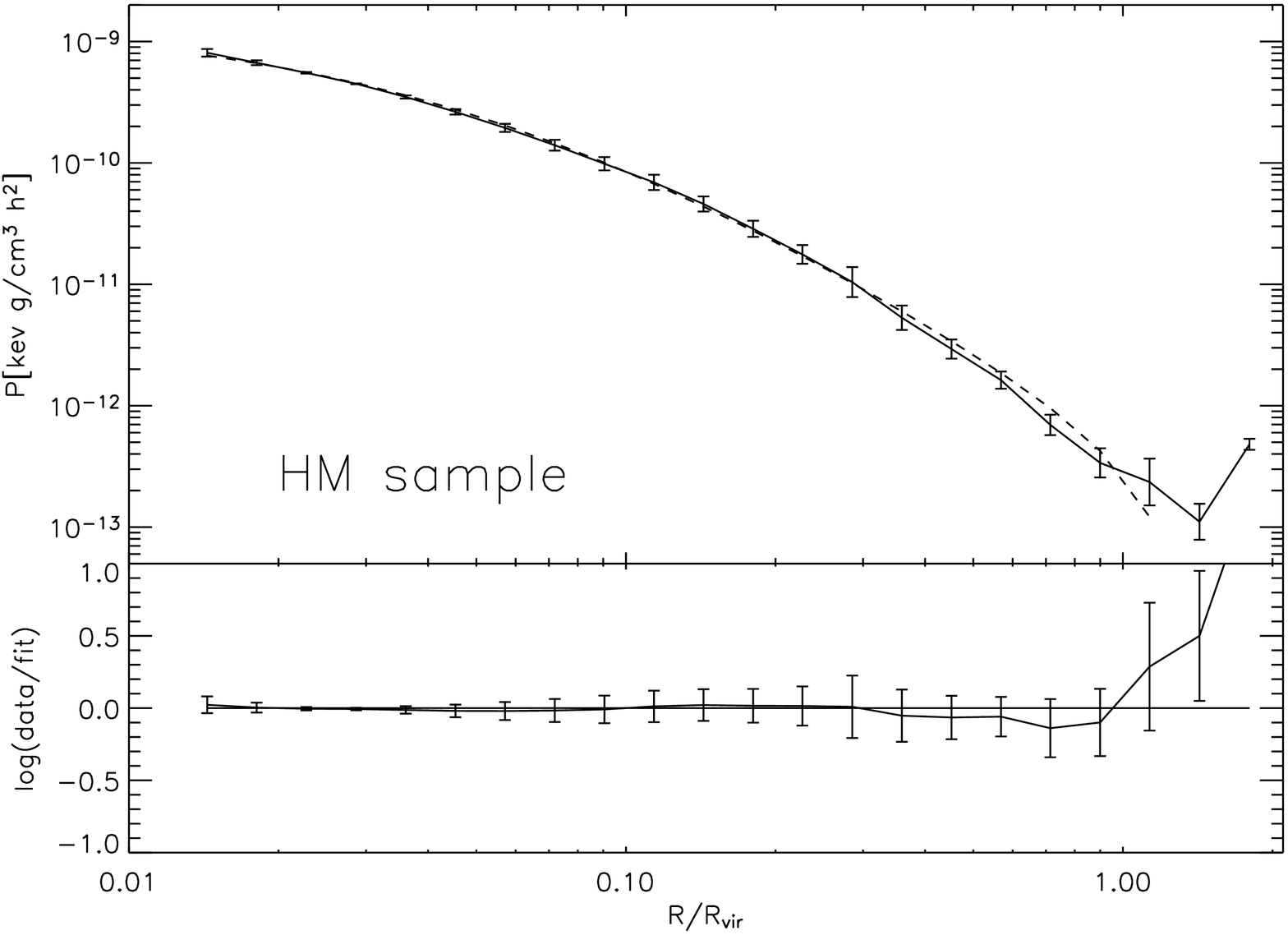} 
\includegraphics[width=8cm,height=7.3cm,angle=90]{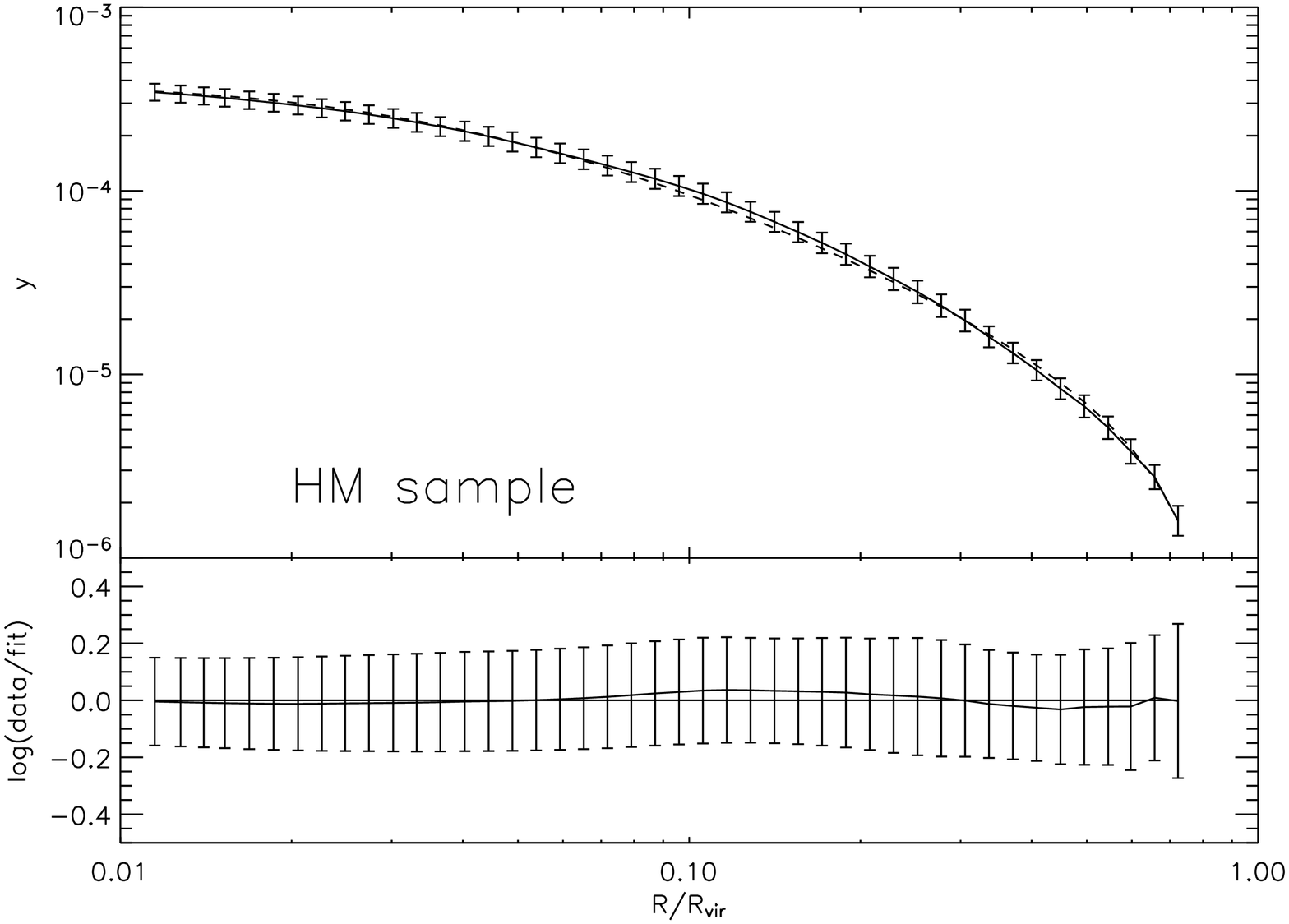} 
\caption{
The radial profiles for 3D pressure $P$ (left panels) and for the
Comptonization parameter $y$ (right panels) as obtained from the
\emph{ovisc} simulations. Upper and lower panels refer to the LM 
and HM samples, respectively. Solid and dashed lines are the mean 
of the logarithmic profiles and our fitting relations, respectively.  
Error bars show error on the mean.  The smaller box below each panel 
shows the mean of the logarithmic residuals with the corresponding root
mean square (rms) value.}
\label{fig:uno} 
\end{center} 
\end{figure*}

As an example, in Fig. \ref{fig:uno} we show radial profiles for
the pressure $P$ and for the Comptonization parameter $y$ (left and
right panels, respectively). Results are displayed for the physical
model \emph{ovisc} and for both cluster samples: LM (upper panels) and
HM (lower panels). Solid lines indicate the mean of the logarithmic 
profiles, with corresponding 1$\sigma$ error bars, while dashed lines 
show the fitting relations obtained using Eqs.(\ref{3dfit}) and 
(\ref{2dfit}).  We notice that the 3D fit is not always within the 
error bars: this is mainly due to the noisy tail of the pressure profile 
outside $R_{\rm vir}$.  Our interpretation is confirmed by looking at 
the logarithmic residuals shown in the smaller box below each panel: 
here we plot their mean and rms values.  At any rate, we will see
that SZ profiles are more regular, as a result of the integration
process, and the integrated relation will be an extremely good fit.

Extending our analysis to the simulations which include different 
physical processes, we computed the fitting parameters which let us
derive the 3D fitting profiles according to Eq.(\ref{3dfit}), and
the 2D ones according to Eq.(\ref{2dfit}). The results are summarized
in Table \ref{tab:due}.
 
\begin{table*} 
\caption{Fitting parameters for 3D and 2D profiles (Eqs.(\ref{3dfit})
and  (\ref{2dfit}), respectively).} 
\label{tab:due} 
\begin{tabular}{|c|c|c|c|c|c|c|} 
\hline 
Cluster sample&Physics&$I_{3D}$[erg/cm$^3$] & $I_{2D}$& $r_p$&
$b$& $K$\\
\hline 
LM sample &\emph{ovisc}&$7.48\times 10^{-14}$&$1.83
\times 10^{-7}$&0.045&-2.75&0.0\\* 
&\emph{lvisc}&$8.93\times 10^{-14}$&$1.88 \times 10^{-7}$&0.075&-2.9&0.0\\ 
&\emph{csf}&$1.00\times 10^{-13}$&$2.36\times 10^{-7}$&0.06&-2.55&0.0\\* 
&\emph{csfc}&$8.10\times 10^{-14}$&$1.99\times 10^{-7}$&0.06&-2.55&0.0\\ 
\hline 
HM sample&\emph{ovisc}&$6.37\times 10^{-13}$&$ 
3.94\times 10^{-6}$&0.05&-2.6&-0.45\\* 
&\emph{lvisc}&$6.56\times 10^{-13}$&$5.26\times 10^{-6}$&0.08&-2.6&-0.45\\ 
&\emph{csf}&$6.56\times 10^{-13}$&$3.52\times 10^{-6}$&0.09&-2.8&-0.45\\* 
&\emph{csfc}&$6.43\times 10^{-13}$&$3.74\times 10^{-6}$&0.09&-2.8&-0.45\\ 
\hline 
\end{tabular} 
\end{table*} 
 
\begin{figure} 
\begin{center} 
\includegraphics[width=0.35\textwidth,angle=90]{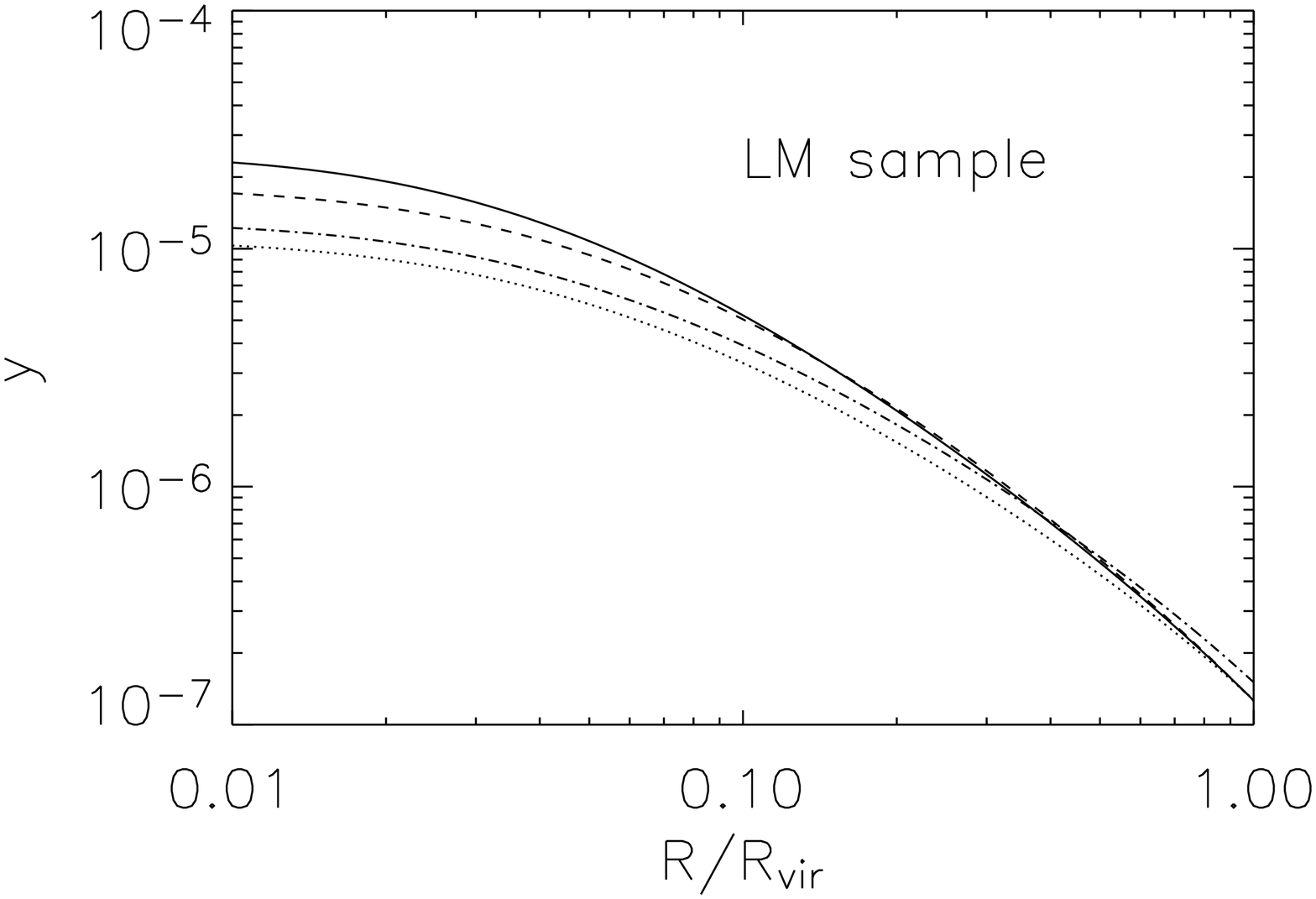} 
\includegraphics[width=0.35\textwidth,angle=90]{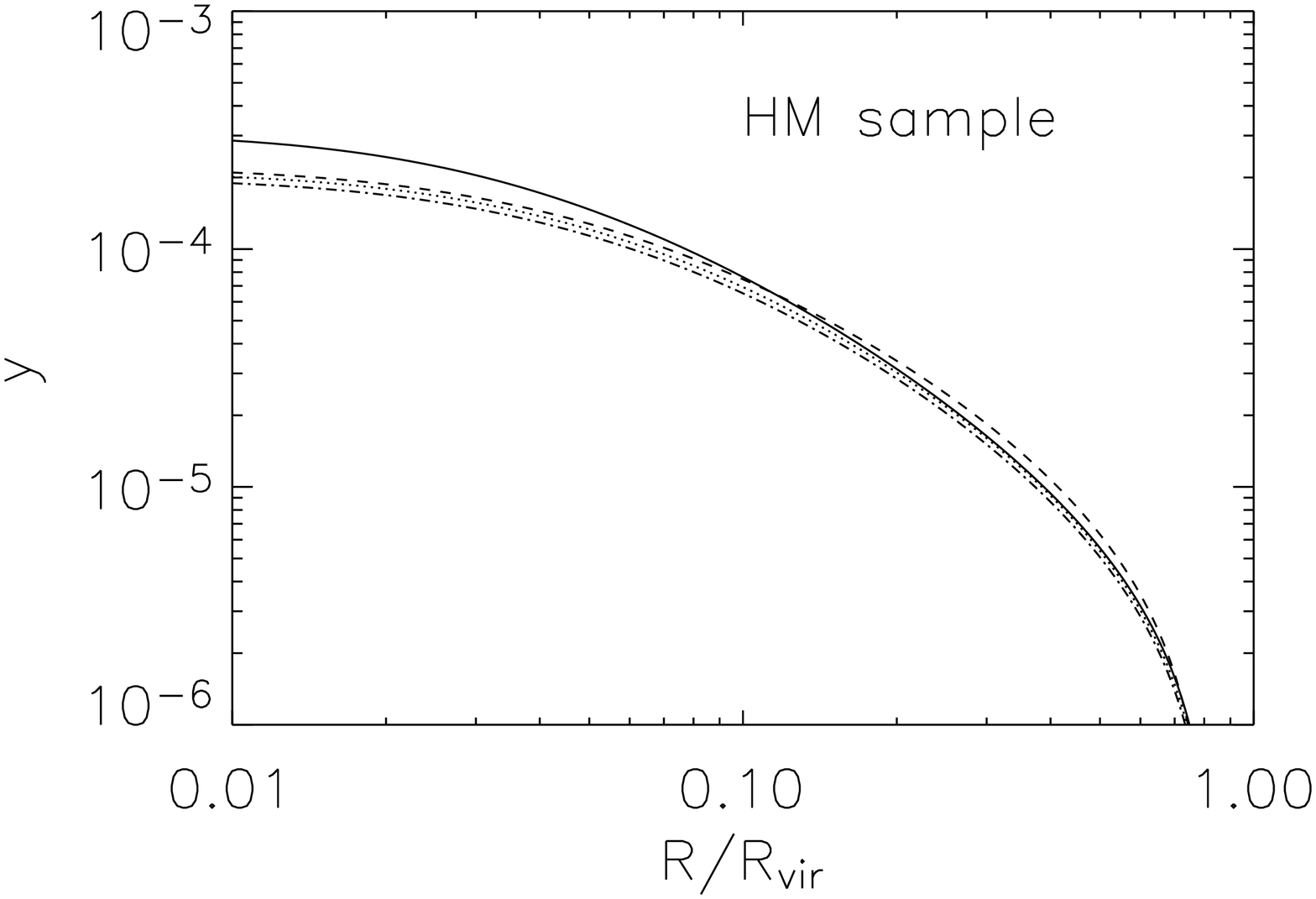} 
\caption{
Comparison between the mean logarithmic SZ profiles obtained from
cluster simulations including  different ICM modelling:
\emph{ovisc} (solid line), \emph{lvisc} (dashed line), \emph{csf}
(dashed-dotted), \emph{csfc} (dotted line).  Upper and lower panels
refer to LM and HM samples, respectively.}
\label{fig:tre} 
\end{center} 
\end{figure}

The resulting SZ profiles for all physics are plotted in Fig.\ref{fig:tre}, 
for both mass bins.  We find that differences due to cluster physics are 
mainly visible in the low-mass clusters, and affect in particular the inner 
region of the profiles: therefore such differences are unlikely to be probed 
through observations (see the discussion below). We also note that, for 
both mass bins, the models \emph{csf} and \emph{csfc} have the same $r_p$, $b$,
and $K$, and only slightly differ in intensity $I_{2D}$. This is due to 
the fact that thermal conduction acts mainly locally, so it does not 
significantly affect quantities averaged on sufficiently large scales, 
like the radial SZ profile.
 
As we stated above, there is a significant difference between the mean
profiles obtained for high- and low-mass clusters, as clearly shown in
Fig. \ref{fig:tre}. We note that there is also a selection
effect which may be responsible for this effect: the HM clusters of our
re-simulations, infact, have been chosen to be relaxed and isolated,
while the LM clusters have more varied dynamical configurations.  As
also pointed out by \cite{rasia2004}, the density profiles in the
external regions are on average steeper for isolated systems than for
objects having more general environmental situations.  This property,
then, reflects in our SZ profiles.  Specifically, profiles of both
subsamples have a logarithmic slope going to 0 as $R/R_{\rm
vir}\rightarrow 0$. The slope of the LM sample steepens at $R/R_{\rm vir}
= 0.04-0.07$ and then remains nearly constant up to $R_{\rm vir}$.
Conversely, the slope of the profiles in the HM sample presents a
first steepening at $R/R_{\rm vir} = 0.05-0.09$ and a second one at
$R/R_{\rm vir} \sim 0.6$.  To reproduce this feature we had to add a
negative constant ($K$) to the fitting relations.

\begin{figure} 
\begin{center} 
\includegraphics[width=0.65\textwidth,angle=90]{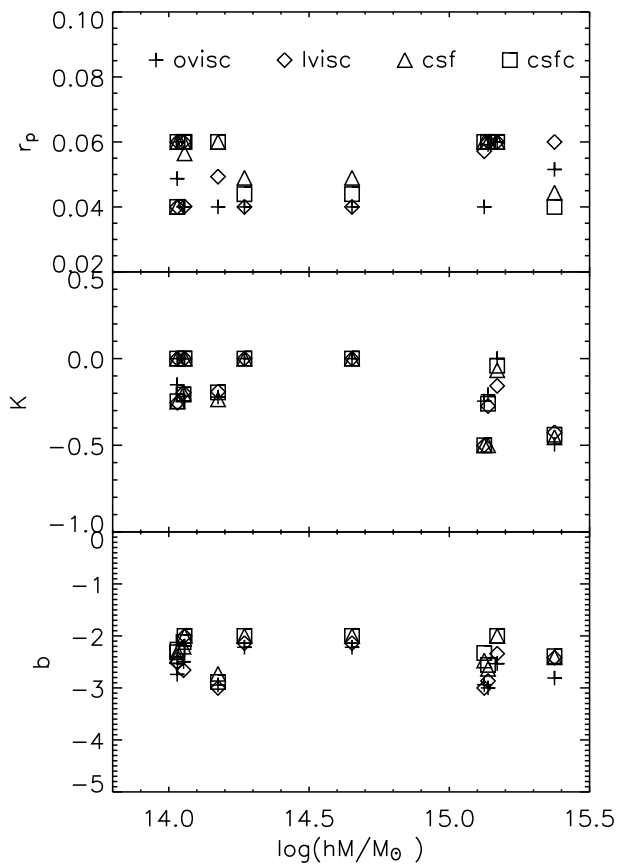} 
\caption{
Mass dependence of the fitting parameters $r_p$ (upper panel), $K$
(central panel) and $b$ (lower panel).  Different symbols refer to
simulations with different physical processes included, as indicated
in the upper panel.}
\label{fig:massdep} 
\end{center} 
\end{figure} 

This result suggests that the fitting parameters $r_p$, $b$ and $K$
might depend on the cluster mass. To investigate whether this is true, 
we fitted relations like Eq.(\ref{2dfit}) to all individual SZ profiles, 
and plotted in Fig. \ref{fig:massdep} the resulting parameters vs. cluster 
mass. Since the mean profiles depend only mildly on the cluster physics, 
we prefer to show the data coming from the different re-simulations 
together, thus increasing our statistics.  Plotting the parameters 
$r_p$, $K$ and $b$ against the (logarithm of the) mass we do not find 
any statistically significant trend.  However, we notice that, as 
suggested by the comparison of the mean profiles, higher masses
correspond on average to higher values for the scale radius $r_p$ and
to steeper profiles at $R/R_{\rm vir} \rightarrow 1$, as shown by the
higher (absolute) value assumed by $K$.

%##############################################################################
%################### SZ Scaling relations######################################
%##############################################################################
 
\section{SZ Scaling relations}\label{sec:quattro} 
 
\begin{figure} 
\begin{center} 
\includegraphics[width=0.65\textwidth,angle=90]{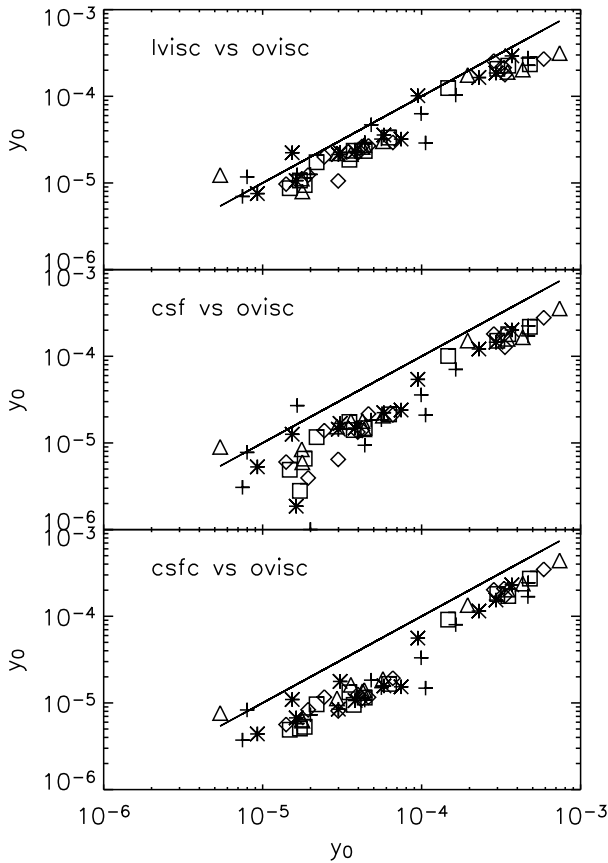} 
\caption{
Comparison of the values $y_0$ obtained for a given cluster in the 
\emph{ovisc} simulation (horizontal axis) with the corresponding 
values obtained in the other simulations (vertical axes): \emph{lvisc} 
(top panel), \emph{csf} (central panel) and \emph{csfc} (bottom panel).  
Different symbols refer to clusters observed at different redshifts: 
$z=0$ (crosses), $z=0.2$ (stars), $z=0.4$ (squares), $z=0.7$ (diamonds) 
and $z=1$ (triangles). The solid line corresponds to the bisecting line.  }
\label{fig:nove} 
\end{center} 
\end{figure} 
 
\begin{figure} 
\begin{center} 
\includegraphics[width=0.65\textwidth,angle=90]{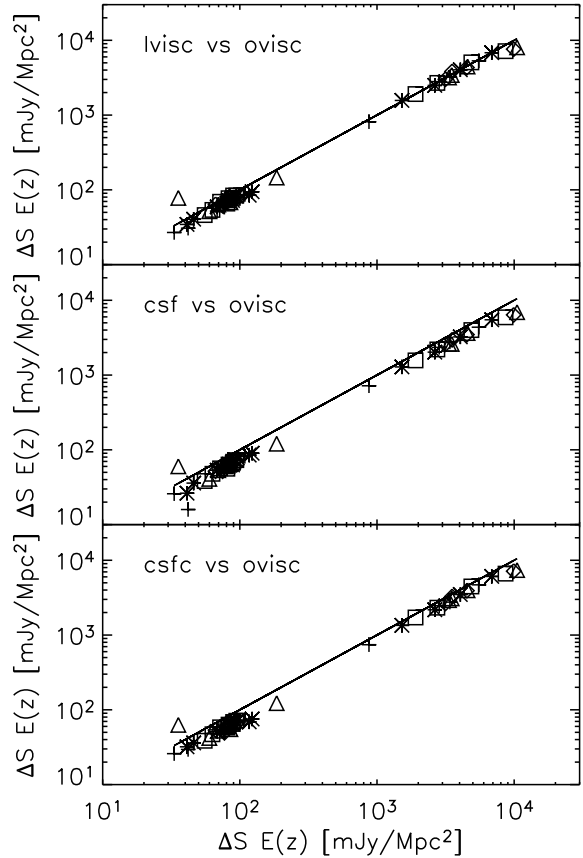} 
\caption{As Fig.\ref{fig:nove}, but for $\Delta S$.}  
\label{fig:dieci} 
\end{center} 
\end{figure} 

If gravity is the only physical process in action during their formation, 
galaxy clusters can be modelled using the so-called self-similar model 
\citep{kaiser1986,navarro1995}, which predicts simple scaling relations 
between the main cluster properties (mass, temperature, etc.). 
In particular, in terms of SZ and X-ray observables, the self-similar 
model suggests the following relations:
 
\begin{eqnarray} 
&y_0& \propto L^{3/4}_X E^{1/4}(z)\ , 
\label{eq:y0L} \\ 
&y_0& \propto T^{3/2} E(z)\ ,  
\label{eq:yoT} \\
&\Delta S& \propto T^{5/2} E^{-1}(z)\ .
\label{eq:deltaT}
\end{eqnarray} 
 
In the previous equations $y_0$ is the central value for the Comptonization 
parameter $y$, $\Delta S$ is the SZ flux integrated out to a given radius, 
$T$ is the ICM temperature, $L_X$ is the X-ray luminosity.  The factor $E(z)$, 
representing the ratio between the Hubble constant at redshift $z$ and its 
present value, is given, for flat cosmologies, by
\begin{equation} 
E^2(z)=\Omega _{\rm 0m} (1+z)^3+\Omega _{\Lambda} \ ,   
\end{equation} 
and allows rescaling all the data to the same redshift (usually $z=0$).

The SZ self-similar scaling relations have been largely investigated
with semi-analytic models \citep[see,
e.g.,][]{cavaliere2001,verde2002,mccarthy2003} and in numerical hydrodynamic
simulations \citep[see, e.g.,][]{dasilva2004,diaferio2005,motl2005},
since deviations from them would indicate that non-gravitational
processes, like gas cooling and energy feedback, play an important
role in gas-dynamics. It is worth noticing that strong support in
favour of a non-negligible role of the non-gravitational processes
already comes from the study of self-similar relations in the X-ray
band \citep[see, e.g.][ and references
therein]{cavaliere1998,allen1998,markevitch1998,xue2000,
borgani2004,arnaud2005,vikhlinin2006}.

We studied the relations in Eqs.(\ref{eq:y0L})-(\ref{eq:deltaT}) by
analysing the same cluster sample of the previous section, but
considering each cluster at five different redshifts: $z=1$, $z=0.7$,
$z=0.4$, $z=0.2$, $z=0$.  This corresponds to a time separations $\sim
2$ Gyr, long enough to allow changes in the cluster dynamical status,
and thus to make successive data points from the same cluster roughly
independent on each other.

\begin{table}
\caption{
Best-fitting parameters for the different SZ scaling relations
($y_0-L_X$, $y_0-T$ and $\Delta S-T$) obtained from our
hydrodynamic simulations with different physical modelling of ICM. 
X-ray luminosity in the [0.1-10 keV] band $L_X$, mass-weighted
temperature $T$ and integrated SZ flux $\Delta S$ are given in
units of $10^{44}$ erg/s, keV and mJy/Mpc$^2$, respectively.}
\label{tab:sei}
\begin{tabular}{|c|c|c|}
\hline
Physics&  $y_0-L_X$ [10$^{44}$ erg/s] \\
\hline
\emph{ovisc}&$\log ({y_0}) = (-4.77 \pm 0.01)+(0.78 \pm 0.02)\log (L_X)$\\
\emph{lvisc}&$\log ({y_0}) = (-4.76 \pm 0.01) +(0.77 \pm 0.02)\log (L_X)$\\
\emph{csf}&$\log ({y_0}) = (-4.63 \pm 0.01)+(0.79 \pm 0.02)\log (L_X)$\\
\emph{csfc}&$\log ({y_0}) = (-4.59 \pm 0.01)+(0.81 \pm 0.03)\log (L_X)$\\%
\hline
Physics &  $y_0-T$ [keV] \\
\hline
\emph{ovisc}&$\log ({y_0}) = (-4.70 \pm 0.03)+(1.54 \pm 0.07)\log (T)$\\%
\emph{lvisc}&$\log ({y_0}) = (-4.80 \pm 0.03)+(1.42 \pm 0.06)\log (T)$\\
\emph{csf}&$\log ({y_0}) = (-5.11 \pm 0.03) +(1.71 \pm 0.05)\log (T)$\\
\emph{csfc}&$\log ({y_0}) = (-5.16 \pm 0.01)+(1.75\pm 0.03)\log (T)$\\
\hline
Physics& $\Delta S$ [mJy/Mpc$^2$] $- T$ [keV] \\
\hline
\emph{ovisc}&$\log (\Delta S) = (1.68 \pm 0.01)+(2.40 \pm 0.01)\log (T)$\\%
\emph{lvisc}&$\log (\Delta S) = (1.71 \pm 0.02)+(2.37 \pm 0.02)\log (T)$\\%
\emph{csf}&$\log (\Delta S) = (1.51 \pm 0.01)+(2.50 \pm 0.01)\log (T)$\\%
\emph{csfc}&$\log (\Delta S) = (1.53 \pm 0.01)+(2.46 \pm 0.02)\log (T)$\\%
\hline
\end{tabular}
\end{table}

\begin{figure*} 
\begin{center} 
\includegraphics[width=0.7\textwidth,angle=90]{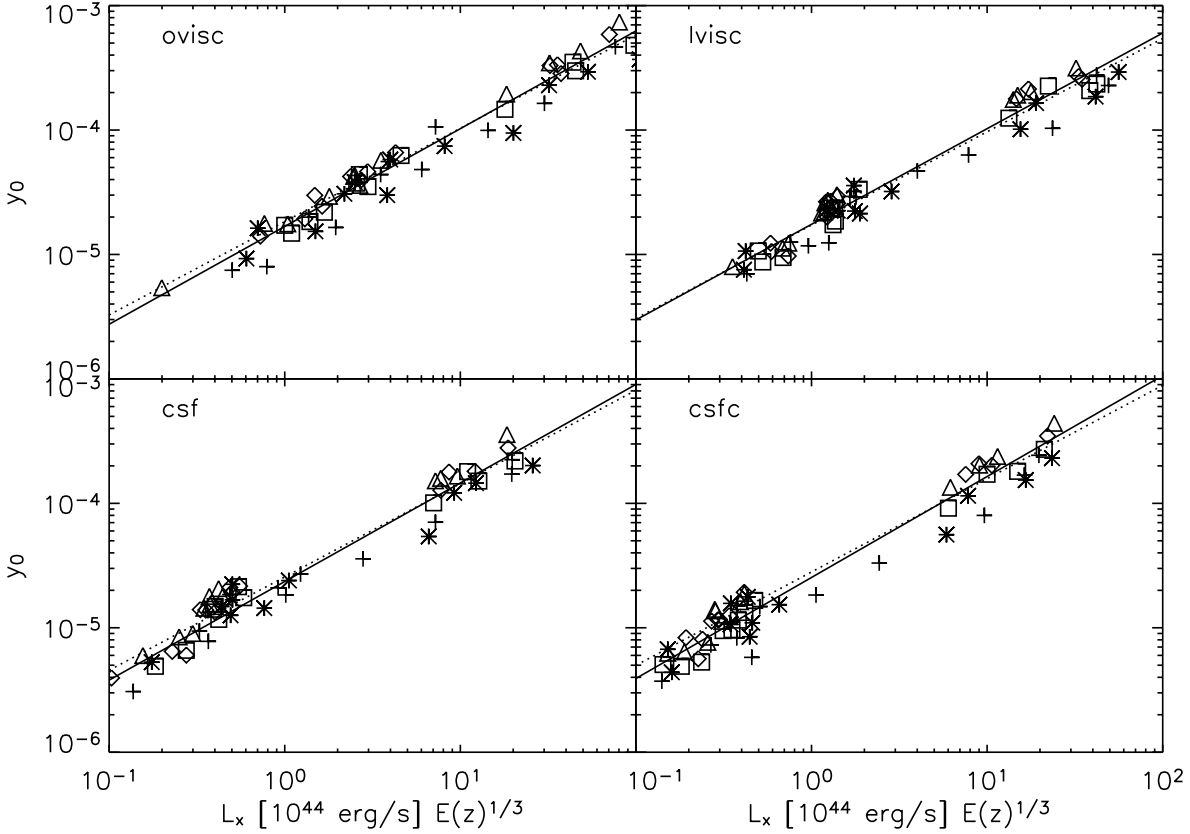} 
\caption{
Scaling relation $y_0-L_X$ for all physics.  Solid and dotted lines
refer to our best-fit relation and to the best fit obtained by
imposing the self-similar slope, respectively.  Different symbols
refer to clusters at different redshifts, as in Fig. \ref{fig:nove}.
}
\label{fig:scalyl} 
\end{center} 
\end{figure*} 

To construct our empirical relations, we used the intrinsic parameters
extracted from our SZ maps.  The X-ray luminosity in the [0.1-10 keV]
band $L_X$ and the integrated SZ flux $\Delta S$ were computed
within $R_{500}$. To estimate the temperature we preferred to adopt a
mass-weighted estimator: this quantity is in fact related to the
energetics involved in the process of structure formation.  As shown
in earlier papers
\citep{mathiesen2001,mazzotta2004,gardini2004,rasia2005}, the 
alternative of using an emission-weighted temperature - originally 
proposed to make results from hydrodynamical simulations directly 
comparable to observational spectroscopic measurements - actually 
introduces systematic biases when the ICM structure is thermally complex.

First we wish to investigate how the SZ signal is modified by the
inclusion of different physical processes.  In Fig. \ref{fig:nove} we
make a one-to-one comparison of the value $y_0$ obtained for each cluster 
at a given redshift in the \emph{ovisc} runs versus those obtained in runs 
including more physics. A similar plot, but for $\Delta S$, is presented 
in Fig.\ref{fig:dieci}.  Different symbols refer to different redshifts, as 
indicated in the figure captions.  We find that the values measured in
the \emph{csfc} and \emph{csf} runs are systematically lower than for
\emph{ovisc}, by a factor $\sim 60$ per cent for $y_0$ and of $25$ per cent 
for $\Delta S$. On the other hand, we tested that the plotted data can be 
fitted with a logarithmic slope which is always compatible with unity,
showing that the slope of the relations is robust against changes in
the modelled physics.

\begin{figure*} 
\begin{center} 
\includegraphics[width=0.7\textwidth,angle=90]{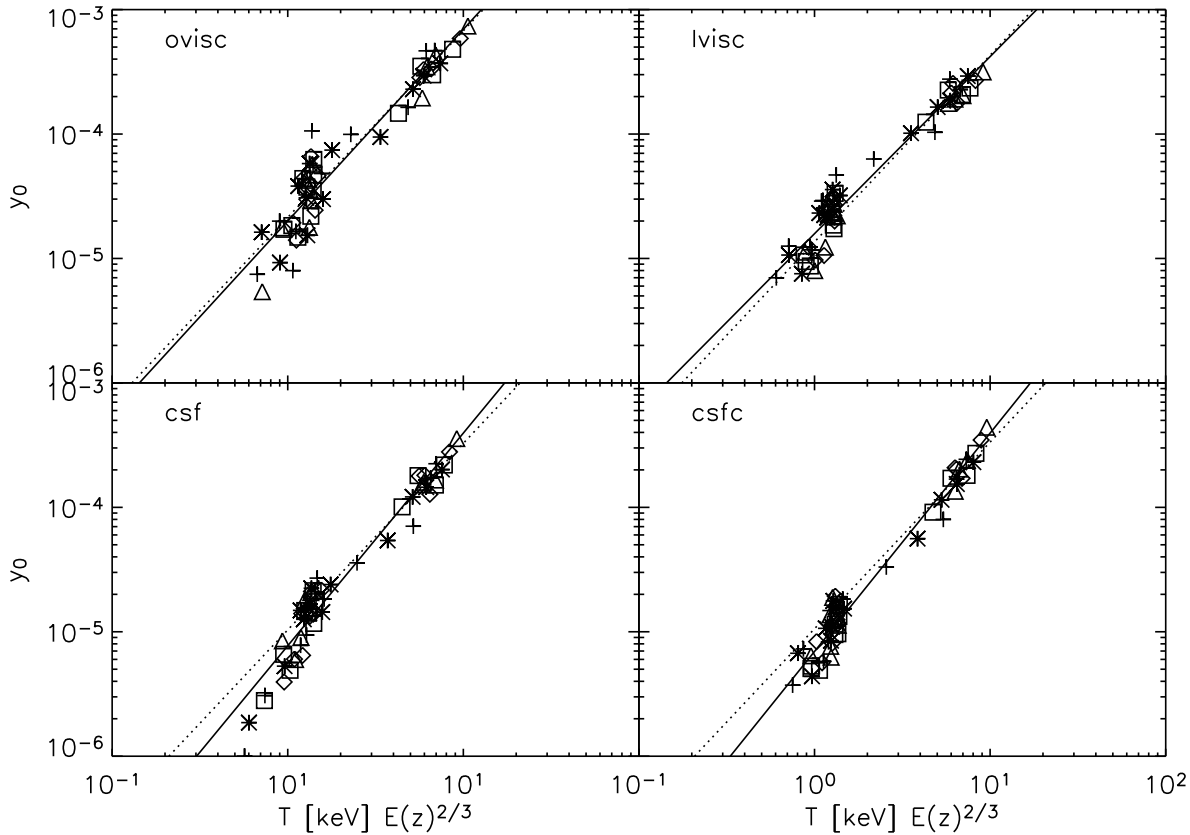} 
\caption{
As Fig.\ref{fig:scalyl}, but for the scaling relation 
$y_0-T$.}
\label{fig:scalyt} 
\end{center} 
\end{figure*} 

Let us now turn to the SZ scaling relations.  The slope and the
normalisation of our best fit relations, with the corresponding errors,
have been obtained by applying to each set of parameters $(x,y)$ a
``robust" least absolute deviation method. Note that here $x$ and $y$
actually express a log-log relation.  First, we fitted the data assuming 
a linear model $y = A_1 + B_1 x$; we then repeated the linear fitting 
swapping $x$ and $y$, and obtained the new parameters ($A_2$ and $B_2$).  
Finally, the slope and intercept $(A,B)$ and the corresponding errors 
$(\Delta A,\Delta B)$ were calculated as follows:
\begin{eqnarray} 
A=(A_1+A_2)/2 &,& \Delta A=|A_1-A_2|/2\\ 
B=(B_1+B_2)/2 &,& \Delta B=|B_1-B_2|/2. 
\end{eqnarray} 

\begin{figure*} 
\begin{center} 
\includegraphics[width=0.7\textwidth,angle=90]{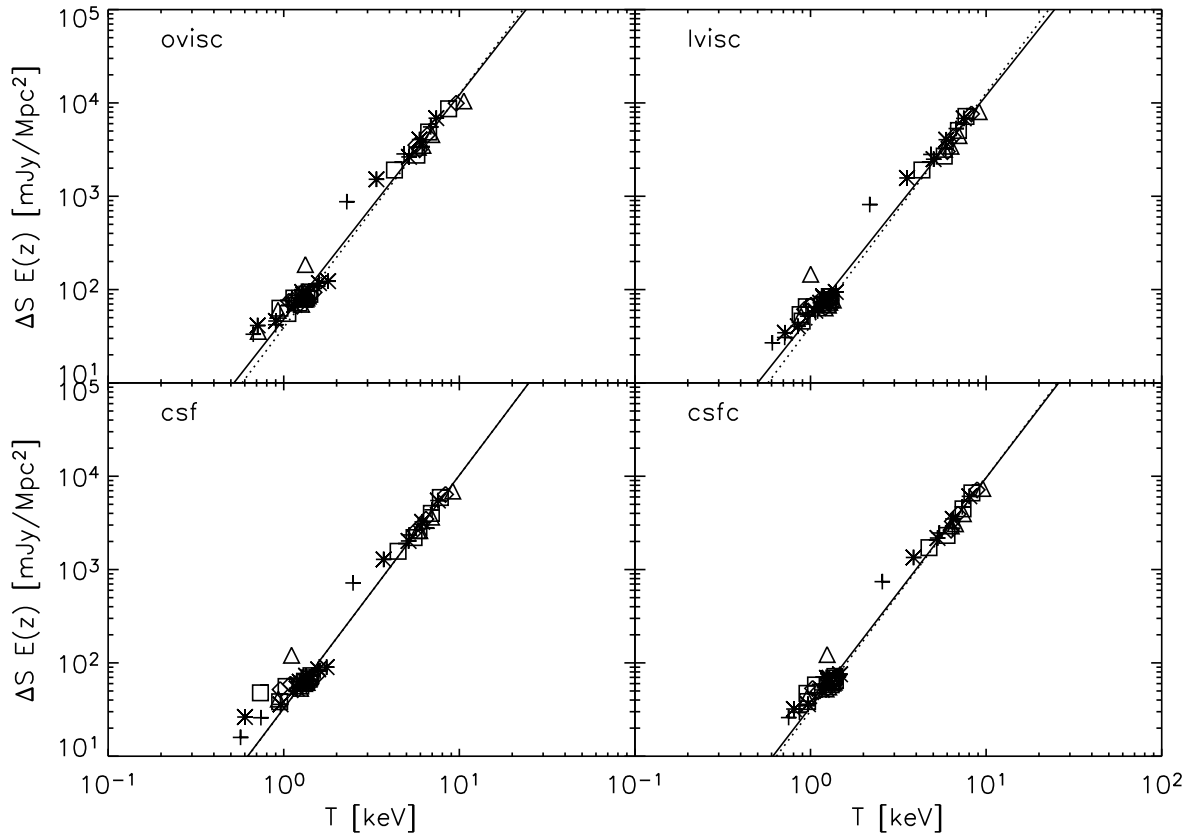} 
\caption{As Fig.\ref{fig:scalyl}, but for the scaling relation
$\Delta S-T$.  }
\label{fig:scalst} 
\end{center} 
\end{figure*}

Our best fit relations, reported in Table \ref{tab:sei}, are also
plotted as solid lines in Figs. \ref{fig:scalyl}, \ref{fig:scalyt} and
\ref{fig:scalst} for the scaling relations between $y_0$ and $L_X$,
$y_0$ and $T$, and $\Delta S$ and $T$, respectively.
For comparison, in each plot the dotted line is the best fit when the
self-similar value for the slope is imposed.  We first notice that data 
belonging to different redshifts and corrected by the appropriate factor 
$E(z)$ define a tight relation: this indicates that the evolution in 
redshift is in good agreement with predictions of the self-similar model, 
at least up to $z=1$.  Since our scaling relations were obtained using a
sufficiently large number of points, the errors associated to the
best fitting slopes and normalizations are relatively small: this increases
the statistical significance of our results.

The empirical $y_0-L_X$ relation for all physics is consistently slightly 
steeper than the self-similar slope, 0.75; the maximum difference is 
$\sim$ 0.05, coming from simulations which include radiative processes. 
The recovered $y_0-T$ relation is shallower than the self-similar
relation for \emph{lvisc}, while for all other physics it is steeper
by $0.1-0.2$.  Finally, the $\Delta S-T$ relation is very close to the
self-similar one (having slope 2.5) for the \emph{csf} runs, but
shallower (by 0.05-0.1) for the other ICM models; this conflicts
with the work by \cite{nagai2005}, who finds indications (though with
no statistical significance) of a slope steeper than the self-similar one.

In general, comparing the values in Table \ref{tab:sei} we find that
the slope of the relations depends very little on the cluster
physics. Similar results were found by \cite{nagai2005}, who reports
that introducing radiative cooling lowers the normalization of the 
$\Delta S-T$ relation, but has no significant impact on the slope. 
In particular, he finds a normalization lower by $34$ per cent, higher 
than our result. Moreover, we notice that also the \emph{lvisc} clusters 
exhibit a normalization lower than the \emph{ovisc} ones: part of the 
pressure is in fact missing, because it was converted into turbulent 
motions. Note that this process is completely different from the pressure
reduction in the runs \emph{csf} and \emph{csfc}, where cooling removes 
gas from the hot phase in the cluster centre.  Of course, it is possible
that in nature both processes are acting at the same time. Interestingly, 
we find that at the centre the signal reduction produced by turbulence 
and by cooling are comparable in high-mass clusters.
 
%##############################################################################
%################### Mock observations ########################################
%##############################################################################
 
\section{Mock observations}\label{sec:cinque} 
 
It is now important to verify if the small signatures produced by the
different physical processes included in the hydrodynamic simulations
discussed above can be detected with a significant signal-to-noise ratio
using current SZ instruments. In order to do that, we produced mock 
observations of our clusters.  Among the many dedicated instruments 
which are planned \citep[see, e.g.,][]{mohr2002,lo2002,kosowsky2003,ruhl2004}, 
we decided to consider the Arcminute MicroKelvin Imager \citep[AMI;][]{jones2002},
which is in its commissioning phase in Cambridge (UK)\footnote{see,
for more detail, {\emph{http://www.mrao.cam.ac.uk/telescopes/ami/}}}
and has already obtained a first detection of a SZ decrement pointing
towards the cluster A1914 \citep{barker2006}.  More in detail, AMI is
a compact array of 10 small (3.7m) dishes packed together and having a
low noise, high-bandwidth back-end system. A complementary array of 8
much larger (13m) dishes will allow the detection and subtraction of
unresolved radio sources.  Both arrays operate at frequencies of 12-18
GHz. The field of view is $21 \times 21$ arcmin.

\begin{figure} 
\begin{center} 
\includegraphics[width=0.35\textwidth,angle=270]{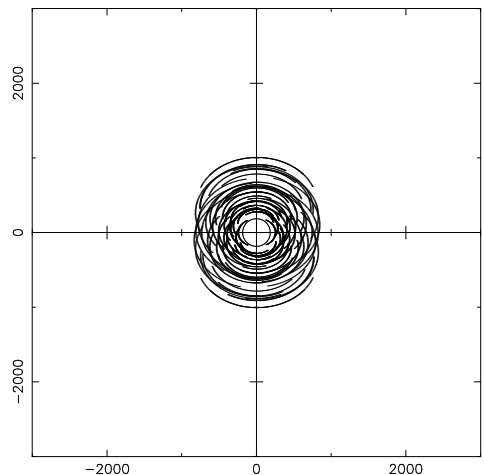} 
\caption{Typical AMI $u,v$ sampling (courtesy of K. Grainge) adopted for
our mock observations. It corresponds to a field at dec $50^\circ$ 
observed at HA=[-4,+4] hours. Axes labels are in units of the wavelength.}
\label{fig:uv}
\end{center} 
\end{figure}

Obtaining a mock observation with the AMI characteristics requires 
simulating the response of the interferometric array to some signal 
on the sky. To do this, we Fourier transformed the image of the simulated 
cluster, and sampled in the $u,v$ plane according to the 
array configuration. The $u,v$ sampling adopted for our observations, and 
shown in Figure \ref{fig:uv}, was kindly provided by the AMI 
collaboration: it corresponds to a field at dec $50^\circ$ and HA [-4, +4] hours. 

We then converted our SZ maps from the Comptonization parameter
$y$ to a measured flux decrement (given in mJy). We subsequently performed 
the sampling in the $u,v$ plane and added noise at the appropriate level. 
Following \cite{kneissl2001}, the noise resulting in the image is 
$\Delta S_{\rm rms}=20$ mJy s$^{-1/2}$ over the field of view here 
considered. In order to reduce noise and spurious structures due to the 
sampling, we smoothed the map and applied the CLEAN deconvolution algorithm 
in order to detect the cluster signal.  The resulting map, calibrated in 
flux for the instrumental efficiency, is what we call ``observed'' cluster.

Notice that in this analysis we do not consider any bias due to CMB 
fluctuations. The spectrum of primary CMB anisotropies 
 may possess significant power on the largest scales probed 
by AMI; however, on scales of few arcmin the SZ signal clearly dominates. 
As we will see, in our case we have good resolution ($\sim 2$ arcmin), and 
we consider only bright objects, so including the effect of the CMB should 
not really change our results.

\begin{figure*} 
\begin{center} 
\includegraphics[width=0.8\textwidth,angle=0]{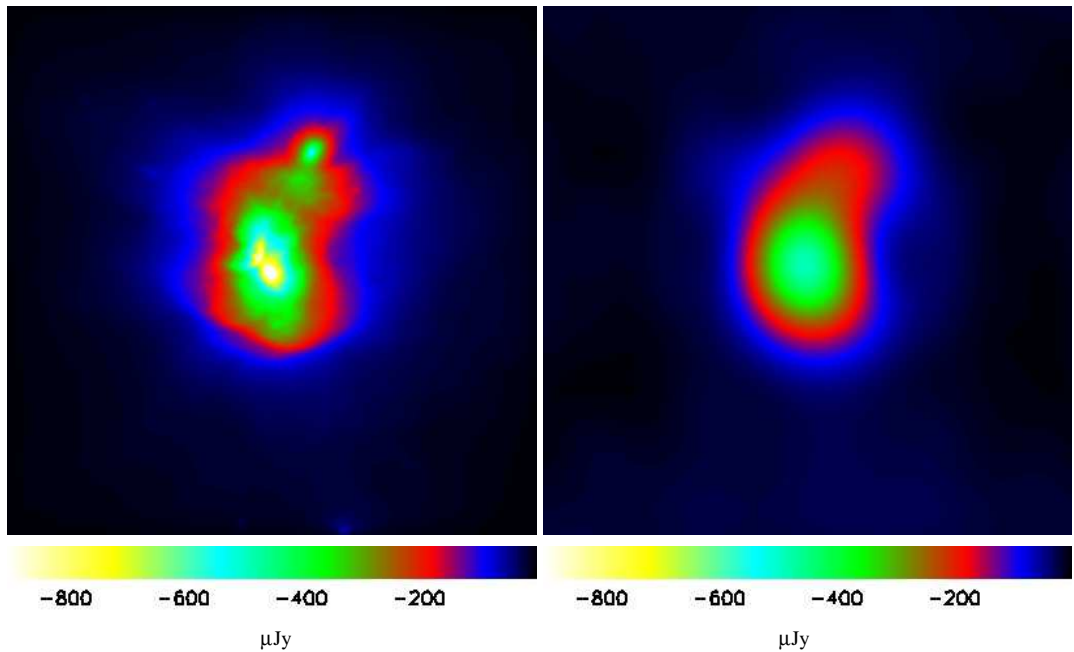}
\caption{Maps for the SZ decrement for the 
\emph{ovisc} simulation of cluster \emph{g72a}. The original map extracted 
from the hydrodynamic simulation, and the same map in the simulated
observation ($t=34h$) which assumes the AMI interferometric response, are
shown in the left and right panel, respectively. The side of each map
corresponds to 16 arcmin. The colour scale is shown at the bottom of
each panel.}
\label{fig:sim_vs_obs} 
\end{center} 
\end{figure*}

For this observational-like approach we considered the simulated
clusters of the HM sample only: in fact, in order to get a signal-to-noise 
ratio ($S/N$) good enough to allow robust extraction of radial
profiles, the exposure time for small-mass objects would have to be too 
high. Moreover, we considered the objects at $z=0.2$: at this redshift our 
clusters cover a significant part of the field, allowing the instrument
to resolve even small structures.
The integration time of each simulated observations is $t=34 h$,
translating into detections at 10 to 20$\sigma$ significance.  
As an example, in Fig.\ref{fig:sim_vs_obs} we compare one map of the SZ 
decrement extracted directly from the hydrodynamic simulation to the
corresponding one obtained at the end of the AMI observational procedure.  
The figure refers to the \emph{ovisc} run of cluster \emph{g72a}, which 
undergoes an important merging event at $z\approx 0.2$.  It is evident
that in the observed map many details are completely lost and isoflux
contours appear more regular.  Moreover, the value of 
SZ decrement at the centre is reduced, as a result of the reduced 
resolution.

\begin{figure} 
\begin{center} 
\includegraphics[width=0.35\textwidth,angle=90]{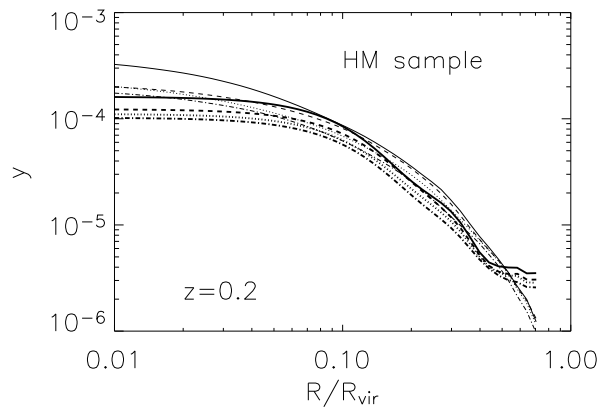} 
\caption{
The mean logarithmic SZ profiles obtained from mock observations of
simulated galaxy clusters are shown as thick lines.  Different
lines refer to simulations including different sets of physical
processes: line types are as in Fig.\ref{fig:tre}. For comparison,
the thin lines refer to the corresponding mean logarithmic SZ profiles
directly extracted from the hydrodynamic simulations.  Only the HM
sample at $z=0.2$ is considered here.}
\label{fig:prof_obs}
\end{center} 
\end{figure}

After preparing the individual observations, we extracted the mean
profiles as described in the previous sections.  The results for
clusters simulated including different ICM modelling are shown by the
thick lines in Fig. \ref{fig:prof_obs}.  In the same plot, for
comparison, we also display as thin lines the mean profiles directly
extracted from the hydrodynamic simulations.  The ripples in the
observed profiles are due to the fact that our simplified removal
procedure could not completely cancel spurious structure introduced by 
the sampling. 

%the negative ring of the dirty beam,
%which cannot be completely cancelled by our simplified removing
%procedure.  
As expected, the observational procedure makes it impossible to
appreciate differences between SZ profiles produced considering different 
physical processes. The observed profile is modified by the instrumental 
beam, so the physics mainly affects the total intensity of the emission.  
However, this quantity also depends on the cluster mass, and breaking the 
degeneracy is difficult.  
%We note that this situation could be improved considering
%higher-resolution observations, which can be obtained with array
%configurations optimized to this purpose.

In Table \ref{tab:tabella11} we discuss the robustness of the parameters 
$y_0$ and $\Delta S$ used in the scaling relations, against the 
observational process.  The central Comptonization parameter $y_0$ is 
clearly suppressed by the degraded angular resolution; the amount of
suppression depends both on the physics included and on the shape of 
the specific cluster, ranging from a factor of 1.5 to almost a factor 
of 2.6 for the \emph{ovisc} run of cluster \emph{g72a}. 
Conversely, the parameter $\Delta S$ is less prone to observational 
effects; moreover, the amount of this effect is very similar for all 
clusters and for all physical processes, being always $\sim$20 per cent. 
We can thus say that, with respect to the observational process, the 
parameter $\Delta S$ is much more stable than $y_0$.  As a consequence, 
the results we found for the $\Delta S-T$ relation should hold even if
a properly observed sample of clusters is used.

\begin{table}
\caption{Ratio between the SZ quantities ($y_0$ and $\Delta S$)
extracted directly from simulated clusters and from their mock observations
($y_{\rm 0obs}$ and $\Delta S_{\rm obs})$.  Columns refer to
the different galaxy clusters belonging to HM sample, while rows show
the ratios for simulations including different sets of physical processes.}

\label{tab:tabella11}
\begin{tabular}{|c|c|c|c|c|}
\hline
 & \emph{g8a}&\emph{g1a}&\emph{g72a}&\emph{g51a}\\
\hline
Physics&&$y_0/y_{\rm 0obs}$\\
\hline
\emph{ovisc}&2.24&    2.49&      2.58&      1.97 \\
\emph{lvisc}&1.73&     1.65&     1.98&      1.45\\
\emph{csf}&1.76&   1.84&     1.93&      1.58\\
\emph{csfc}& 2.16&  1.77&  2.20 &  1.75\\
\hline
Physics&&$\Delta S/\Delta S_{\rm obs}$\\
\hline
\emph{ovisc}& 1.19 &  1.19 &   1.21 &  1.18\\ 
\emph{lvisc}& 1.21 &    1.20&      1.20&      1.22 \\
\emph{csf}&1.18&     1.21&     1.17&     1.20\\
\emph{csfc}& 1.18&  1.21 &      1.21 &   1.21\\
\hline
\end{tabular}
\end{table}

%##############################################################################
%################### Conclusions ##############################################
%##############################################################################

\section{Conclusions}\label{sec:conclu}

We used a sample of simulated galaxy clusters with masses ranging 
from $10^{14}$ to about $2\times 10^{15} h^{-1} M_{\odot}$, re-simulated 
using different sets of physical processes (radiative cooling, star
formation, feedback and thermal conduction) to investigate the
dependence of the SZ emission on the ICM physics. Starting from the
3D pressure information, we obtained for each physics fitting relations 
to the SZ radial profiles. We compared the scaling relations linking 
cluster central and total SZ flux, and cluster X-ray luminosity and 
(mass-weighted) temperature, with those predicted by the self-similar 
model. We then performed mock observations of the most massive systems 
in our sample, in order to check whether our findings survive the
observational procedure. We summarize our main results below.
 
\begin{itemize} 
\item 
The shape of the SZ cluster profile strongly depends on the cluster
mass, as a result of the fact that our high mass clusters, at variance 
with the low mass ones, are all isolated and virialized. In particular, 
while the 3D (logarithmic) pressure profiles of
low-mass clusters (LM sample) can be described by a function which
is nearly flat for $R\rightarrow 0$ and steepening for $R/R_{\rm
vir}\sim 0.05$, the high-mass clusters (HM sample) exhibit an
additional steepening of the profiles at $R/R_{\rm vir}\sim 0.6$.
\item 
The details of the ICM physics have little impact on the SZ profiles;
the largest differences are found near the central regions of low-mass 
systems.
\item 
The scaling relations between cluster SZ emission and cluster X-ray
luminosity and temperature evolve with redshift according to the
self-similar model out to $z=1$; this result is consistent with
previous works \citep{dasilva2004, motl2005, nagai2005}.
\item 
The slopes of the scaling relations for all physics are consistent 
with the predictions of the self-similar model, even if small 
discrepancies can still be found: in particular, our empirical 
relations $y_0-L_X$ and $y_0-T$ are generally slightly steeper 
than the self-similar ones, while the $\Delta S-T$ relation is 
generally slightly shallower.
\item 
The ICM physics has little effect on the slopes of the scaling
relations, in agreement with previous studies \citep{white2002,nagai2005}.
\item 
The main effect introduced in the scaling relations by radiative cooling 
is a lower normalization. If we compare our simulations including cooling
with the non-radiative runs with standard viscosity (\emph{ovisc}), we
find a suppression of $50$ per cent in the central intensity $y_0$ and
of $25$ per cent in the integrated flux $\Delta S$. If we compare the
cooling plus star-formation runs with the non-radiative ones adopting
modified artificial viscosity (\emph{lvisc}), these reductions are of
the order of $20-30$ per cent. A similar normalization lowering for the 
$\Delta S-T$ relation was also found by \cite{nagai2005}.  
\item 
Our mock observations suggest that arcminute resolution data from the current generation of SZE imaging experiments will not be a sensitive probe of the different physics that is ongoing in the intracluster medium.  At arcminute scale angular resolution these effects tend to only have observable consequences on the total cluster flux, which leads to a degeneracy between physics and cluster mass that will be difficult to break
\item 
The integrated flux $\Delta S$ is found to be stable against
the observational process, and thus should be preferred to $y_0$, 
which instead exhibits large variations after performing the mock observations. 
\end{itemize} 

%##############################################################################
%###################### Acknowledgements / Bibliography #######################
%##############################################################################

\section*{acknowledgements}

We thank E. Rasia, M. Massardi, P. Mazzotta, S. Borgani and R. Kneissl
for useful discussions and the anonymous referee for her/his comments
which allowed to improve the presentation of our results. We
aknowledge K. Grainge and the AMI collaboration for providing all the
necessary information about the AMI interferometer.  The simulations
have been performed using the IBM-SP4/5 machine at the ``Consorzio
Interuniversitario del Nord-Est per il Calcolo Elettronico'' (CINECA,
Bologna), with CPU time assigned thanks to the INAF--CINECA grant, the
IBM-SP3 machine at the Italian Centre of Excellence ``Science and
Applications of Advanced Computational Paradigms'' (Padova) and the
IBM-SP4 machine at the ``Rechenzentrum der Max-Planck-Gesellschaft''
at the ``Max-Planck-Institut f\"ur Plasmaphysik'' with CPU time
assigned to the ``Max-Planck-Institut f\"ur Astrophysik''. KD
acknowledges partial support by a Marie Curie Fellowship of the
European Community program "Human Potential" under contract number
MCFI-2001-01227.

\bibliographystyle{mn2e} 
%\bibliography{master}

\label{lastpage}
\end{document}